\documentclass[12pt, twoside]{article}
\usepackage{amsmath}
\usepackage{graphicx}
\usepackage{epsfig}
\usepackage{pstricks,pst-3d,pst-text,pst-node}
\usepackage{txfonts}
\usepackage[numbers]{natbib}
\newpsobject{showgrid}{psgrid}{subgriddiv=1,griddots=10,gridlabels=10pt}
\begin{document}
\title{A DFT+$U$ study of Rh, Nb codoped rutile TiO$_2$}
\author{Kulbir Kaur Ghuman and Chandra Veer Singh* \\
\small Department of Materials Science and Engineering, University of
Toronto, 184 College Street,\\ \small Suite 140, Toronto, Ontario M5S 3E4 Canada.\\}
\date{\today}
\maketitle
\begin{abstract}
A systematic study of electronic structure and band gap states is conducted to analyze the mono doping and charge compensated codoping of rutile TiO$_2$ with Rh and Nb, using the DFT+$U$ approach. Doping of rutile TiO$_2$ with Rh atom induces hybridized O 2p and Rh 4d band gap states leading to a red shift of the optical absorption edge, consistent with previous experimental studies. Since Rh mono-doping may induce recombination centers, charge-compensated codoping with Rh and Nb is also explored. This codoping induces an electron transfer from Nb induced states to Rh 4d states which suppresses the formation of Rh$^{4+}$, thereby leading to a reduction in recombination centers and to the formation of more stable Rh$^{3+}$. A combination of band gap reduction by 0.5 eV and the elimination of band gap states that account for recombination centers makes (Rh,Nb)- codoped TiO$_2$ a more efficient and stable photocatalyst. 
\end{abstract}
\maketitle
\section{Introduction}
\par The development of a safe, abundant and inexpensive photocatalytic material is an important consideration in solving our imminent energy and environmental crisis~\cite{Diebold2011}. Titanium dioxide (TiO$_2$) is one such promising material that meets the requirements of a good photocatalyst and has therefore attracted wide attention in recent years for use into solar hydrogen production~\cite{Fujishima2008, Ni2007}, water decontamination~\cite{Ollis1993}, and solar cells~\cite{Wang2001}. However, TiO$_2$ is a wide band gap semiconductor (3.0 eV for rutile and 3.2 eV for anatase). This restricts sunlight absorbance in the UV range, accounting for $<5\%$ of the full solar spectrum and rendering pure TiO$_2$ inefficient for solar energy conversion. Suitable reduction of the band gap, either by doping or by surface engineering, can lead to significant improvements in the photocatalytic activity of TiO$_2$ \cite{Xu2002,Diebold2003,Yang2007,Chen2011a}. Doping leads to defect states in the band gap~\cite{Fuerte2001} which allow sub-band-gap illumination by electronic transitions from VB to defect states or from defect states to CB thereby improving visible light activity. However, these intraband states may also serve as recombination centers leading to a significant drop in the overall photocatalytic efficiency~\cite{Choi1994}.

\par TiO$_2$ doped with rare earth metals possess significantly improved photo activity owing to the enhanced electron density that is imparted to the titania surface by the dopant oxide \cite{Gao2005}. Despite their potential significance, investigations on doping with rare earth metals have so far been limited. In particular, experimental reports of rhodium (Rh) doping suggest that photocatalytic activity depends on the dopant concentration, its distribution, visible light intensity, and the preparation method \cite{Choi1994,Kitano2011}. Choi et al.~\cite{Choi1994} reported that doping TiO$_2$ with Rh increases the photo reactivity markedly, due to the charge transfer transitions between Rh and TiO$_2$ CB or VB; or a d-d transition in the crystal field. Kitano et al.~\cite{ Kitano2011} showed that Rh modified TiO$_2$ samples exhibit higher level of activity than Cu$^{2+}$ modified and the nitrogen-doped samples. On the other hand, Niishiro et al.~\cite{Niishiro2007} reported that TiO$_2$ doped with only Rh shows only negligible photocatalytic activity as it creates Rh$^{4+}$ ions. They suggested that in such cases, charge compensated codoping with pentavalent ions such as Sb$^{5+}$, Nb$^{5+}$, Ta$^{5+}$ becomes necessary for suppressing the formation of Rh$^{4+}$ ions and improving photocatalytic activity. 

\par Motivated by above experimental findings and apparent discrepancy in the role of Rh-dopant, we performed a detailed investigation of the band structure and electronic properties of Rh-doped, Nb-doped  and (Rh,Nb)- codoped rutile. Codoping was investigated in order to understand the potential benefit of charge compensation for eliminating or suppressing the recombination centers that may be present in mono-doped rutile structures. First principles based theoretical investigations were conducted so as to clearly delineate the origin of band gap defect states, the local structural environment around the dopant site and the localization of excess charge due to defects. Furthermore, in order to represent strong d-electrons correlation in titania, the present study utilized the density functional theory with Hubbard energy correction, i.e. the DFT+$U$ methodology. An in-depth comparison of our theoretical findings with above mentioned experimental data is performed so as to ascertain the effect of rare earth metal doping of rutile in the hope of developing an efficient visible light photocatalyst.
\section{Computational details}
Rutile has a tetragonal crystal structure, with a space group of $p42/mnm$ containing two Ti and four O atoms. A 2$\times$2$\times$3 supercell consisting of 72 atoms was considered in the present calculations. The Quantum-Espresso~\cite{QE} software package within the plane-wave pseudopotential approach was utilized throughout. The interaction between the valence electrons and the ionic core was described by the generalized gradient approximation (GGA) in the Perdew-Burke-Ernzerhof (PBE) formulation~\cite{Perdew1996_pbe} with Vanderbilt ultrasoft pseudo-potentials~\cite{Vanderbilt1990_PRB}. The kinetic energy cutoffs of 544 eV (=40 Ry) and 5440 eV were used for wave functions and charge density, respectively. Brillouin zone integrations were performed using a Monkhorst-Pack~\cite{Monkhorst1976} grid of 4 $\times$ 4 $\times$ 6 $k$-points. All calculations were spin polarized. The structures were relaxed using a conjugate gradient minimization algorithm until the magnitude of residual Hellman-Feynman force on each atom was less than 0.025 eV/$\AA$. The bulk lattice parameters for zero pressure rutile structure obtained by variable cell relaxation using DFT, a=b=4.615 $\AA$ and c=2.961 $\AA$, were found to be in good agreement with experiments~\cite{ICSD2010} and other theoretical calculations~\cite{Zeng2010}.

\par Appreciable underestimation of band gap and delocalization of d and f electrons are well known limitations of DFT. Therefore, we used DFT+$U$ formalism in these calculations. Our DFT calculations suggested a band gap value of 1.89 eV, consistent with other theoretical calculations~\cite{Shao2008}. On the other hand, DFT$+U$ approach with $U$ applied to Ti 3d electrons, varying from 1 to 11 eV, revealed a band gap of 2.85 eV at $U=8$ eV, far closer to the experimental value. Nevertheless, choosing $U$ solely on the basis of band gap may lead to incorrect conclusions regarding the effect of dopants on the electronic properties~\cite{Iwaszuk2011}. It must rather be chosen depending on the property of interest~\cite{Deskins2007}, which, in the current study, is the photocatalytic behavior of TiO$_2$ that in-turn depends upon the position of band gap states and their effect on the electronic structure. Spectroscopic experiments for bulk reduced rutile TiO$_2$ crystals suggest a broad gap state lying about 0.9 eV below the CB edge~\cite{Wendt2008, Thomas2007}. Following the approach of Moller et al.~\cite{Moller2010}, it was found that $U$= 4.2 eV yields correct position of band gap states due to Ti interstitial and O vacancy defects, while opening up the band gap to 2.2 eV for rutile. This value of $U$ is consistent with theoretical investigations by Morgan et al.~\cite{Morgan2007}, who calculated it by fitting the peak positions for surface oxygen vacancies to experimental XPS data. Furthermore, the resulting electronic projected density of states (PDOS) plot for Ti interstitial and O vacancy shows that these defect states are composed of mainly titanium d states, in good agreement with experiments~\cite{Yang2010}.
\section{Results and discussion}
Experimentally obtained X-ray diffraction peaks of Rh-doped TiO$_2$ have revealed that Rh is substituted for Ti$^{4+}$ sites~\cite{Niishiro2007}. This is expected because Rh$^{3+}$ ions (0.805 $\AA$) are bigger than Ti$^{4+}$ ions (0.61 $\AA$)~\cite{Shannon1976}. As radius of Nb (0.64 $\AA$) is also comparable to Ti atom, it should also be substituted to Ti atoms. The concentration of Rh and Nb ions in present study is $\sim$1.4$\%$ commensurate with the concentration of dopants considered in experiments~\cite{Niishiro2007}.
\subsection{Structural distortion} 
Doping leads to local structural disorder that modifies the electronic environment and therefore affects photocatalytic properties. Here, structural distortion was investigated by analyzing the local coordination environment, nature of chemical bonding, and dopant-O-dopant distances. Undoped rutile has a symmetric structure where each Ti atom is bonded to four nearest and two second-nearest oxygen neighbors with bond lengths of 1.95~$\AA$ and 1.99~$\AA$, respectively. Introduction of an Rh dopant into rutile leads to the elongation of all six Rh-O distances without breaking symmetry. However, introduction of the Nb dopant leads to elongation of four Nb-O distance and shortening of the other two Nb-O bonds, thereby affecting the symmetry around the dopants to an appreciable extent. A comparison of ionic radii of dopants with respect to Ti atoms suggests that a dopant with smaller ionic radius leads to larger structural distortion. The local structure around the Rh and Nb in codoped TiO$_2$ is strongly distorted as compared to Rh and Nb monodoped TiO$_2$. For (Rh,Nb)- codoped structure the O-Rh bonds lengths are 2.038, 2.063, and 2.029 $\AA$ and O-Nb bond lengths are 2.005, 1.997 and 1.925 $\AA$. Thus, in (Rh,Nb)- codoping four Nb-O distances get elongated, two Nb-O distances get shortened, and all 6 Rh-O distances get elongated. Overall, Nb$^{5+}$ attempts to shorten two cation-O distances and leaves most of the Rh-O distances notably elongated in order to obtain a coordination environment favorable for its ionic radius. This results in a comparatively more distorted structure than mono doped TiO$_2$ structures.  
\subsection{Formation energy}
The formation energy is an important criterion to evaluate the relative difficulty for the incorporation of dopants into the host lattice and therefore  the relative stability of the doped systems. The formation energies of Rh-doping, Nb-doping and (Rh,Nb)- codoping can be written as 
\begin{equation}
 E_f(Rh) = E(Ti_{24-n_1}Rh_{n_1}O_{48})-(E(Ti_{24}O_{48})+n_1\mu_1-n_0\mu_0), \\
\end{equation}
\begin{equation}
 E_f(Nb) = E(Ti_{24-n_2}Nb_{n_2}O_{48})-(E(Ti_{24}O_{48})+n_2\mu_2-n_0\mu_0), \\
\end{equation}
\begin{equation}
 E_f(RhNb) = E(Ti_{24-n_1-n_2}Rh_{n_1}Nb_{n_2}O_{48})-(E(Ti_{24}O_{48})+n_1\mu_1+n_2\mu_2-n_0\mu_0),
\end{equation}
where $E(Ti_{24-n_1}Rh_{n_1}O_{48})$, $E(Ti_{24-n_2}Nb_{n_2}O_{48})$, $E(Ti_{24-n_1-n_2}Rh_{n_1}Nb_{n_2}O_{48})$ and $E(Ti_{24}O_{48})$ are the total energies of the Rh-doped TiO$_2$, Nb-doped TiO$_2$, (Rh,Nb)- codoped TiO$_2$ and pure TiO$_2$, respectively. $\mu_0$, $\mu_1$ and $\mu_2$ are the chemical potentials of the Ti, Rh and Nb elements, respectively; and $n_0$, $n_1$ and $n_2$ denote the numbers of the host (Ti) atoms substituted, Rh-dopant atoms and Nb-dopant atoms, respectively. In the present study, hcp bulk metal Ti, fcc bulk metal Rh and bcc bulk metal Nb were used to determine the chemical potentials. Since the focus here is on the relative stabilities, the choice of chemical potential should not influence the conclusions about stability of the doped TiO$_2$. The calculated formation energies for Rh-doped and Nb-doped rutile were +0.417 eV and -0.0689 eV, respectively. Postive and negative values of formation energy for Rh- and Nb-doped TiO$_2$ respectively, suggest that Nb-doped TiO$_2$ is more stable as compared to Rh-doped TiO$_2$.
\par
\begin{figure*}[]
 \centerline{\hbox{ 
   \epsfxsize=4.0in
   \epsffile{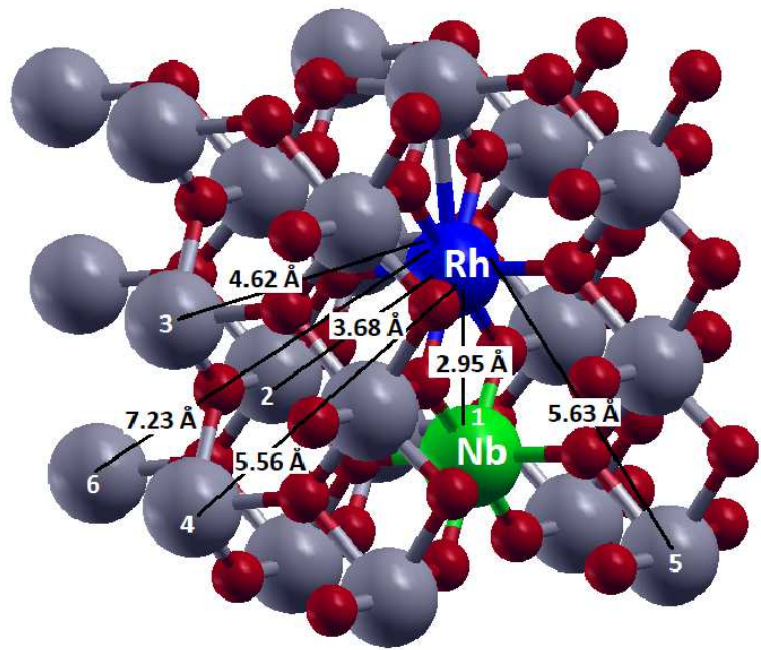}}
  }
 \caption{The (Rh,Nb)- codoped $2\times 2\times 3$ supercell of rutile. Ti, O, Rh and Nb atoms are highlighted in gray, red, blue and green colors, respectively. The six nearest neighbouring positions considered for Nb atom are shown along with dopant-dopant distances.}
  \label{RhNbModel}
\end{figure*}

 Calculation of formation energy is also useful in determining the most stable configuration of the codoped structure.
For this purpose, six nearest-neighbor codoping systems, labeled 1-6 in Fig.~\ref{RhNbModel}, were constructed. The calculated formation energies for Rh-Nb distances of 2.95, 3.68, 4.62, 5.56, 5.63 and 7.23 $\AA$ were respectively equal to 0.232, 0.241, 0.240, 0.239, 0.240 and 0.233 eV. A relatively small range of formation energies of structures with different Rh-Nb distances indicates high stability of the codoped TiO$_2$ structures. For subsequent study of electronic properties, the most stable (Rh,Nb)- codoped structure with dopant-dopant distance of 2.95 $\AA$ and $E_f$=0.232 eV was chosen. The chosen Rh and Nb atoms are highlighted in blue and green respectively, in Fig.~\ref{RhNbModel}. (Rh,Nb)- codoped TiO$_2$ has a smaller formation energy than that of Rh-doped TiO$_2$. Therefore, choosing Nb as codopant not only favors charge compensation but may also lead to spontaneous formation of (Rh,Nb)- codoped system. This finding is consistent with experimental observations~\cite{Niishiro2007}.
On the whole, structural distortion and formation energy results suggest that stability of defect formation depends on the ionic radius of the dopant, in agreement with Nolan~\cite{Nolan2011}. The strong distortion due to small ionic radius of dopant Nb makes (Rh,Nb)- codoped TiO$_2$ more stable than Rh-doped TiO$_2$.
\subsection{Electronic properties} 
In order to understand the photocatalytic properties of doped systems, the spin polarized density of states (DOS) and PDOS were calculated for Rh-doped, Nb-doped and (Rh,Nb)- codoped rutile TiO$_2$. The corresponding spin density plots are also shown. A conventional Gaussian smearing of 0.1 eV was utilized. 
\subsubsection{Rh-doped rutile TiO$_2$}
\begin{figure}
\begin{pspicture}(0,6)(15,24) 
  \rput[tl](0,23){\includegraphics[width=9cm]{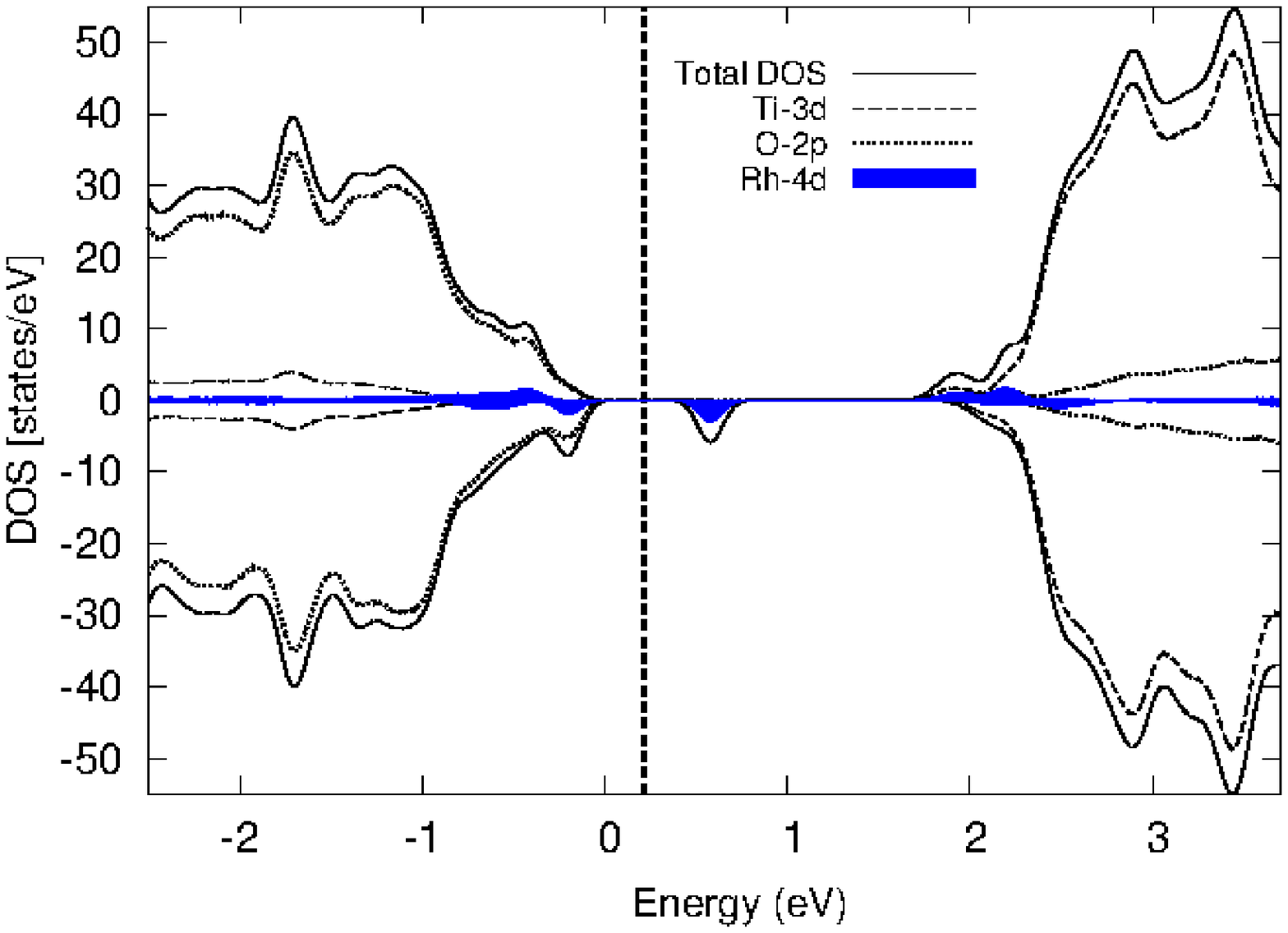}}
  \rput[tl](9,23){\includegraphics[width=6cm]{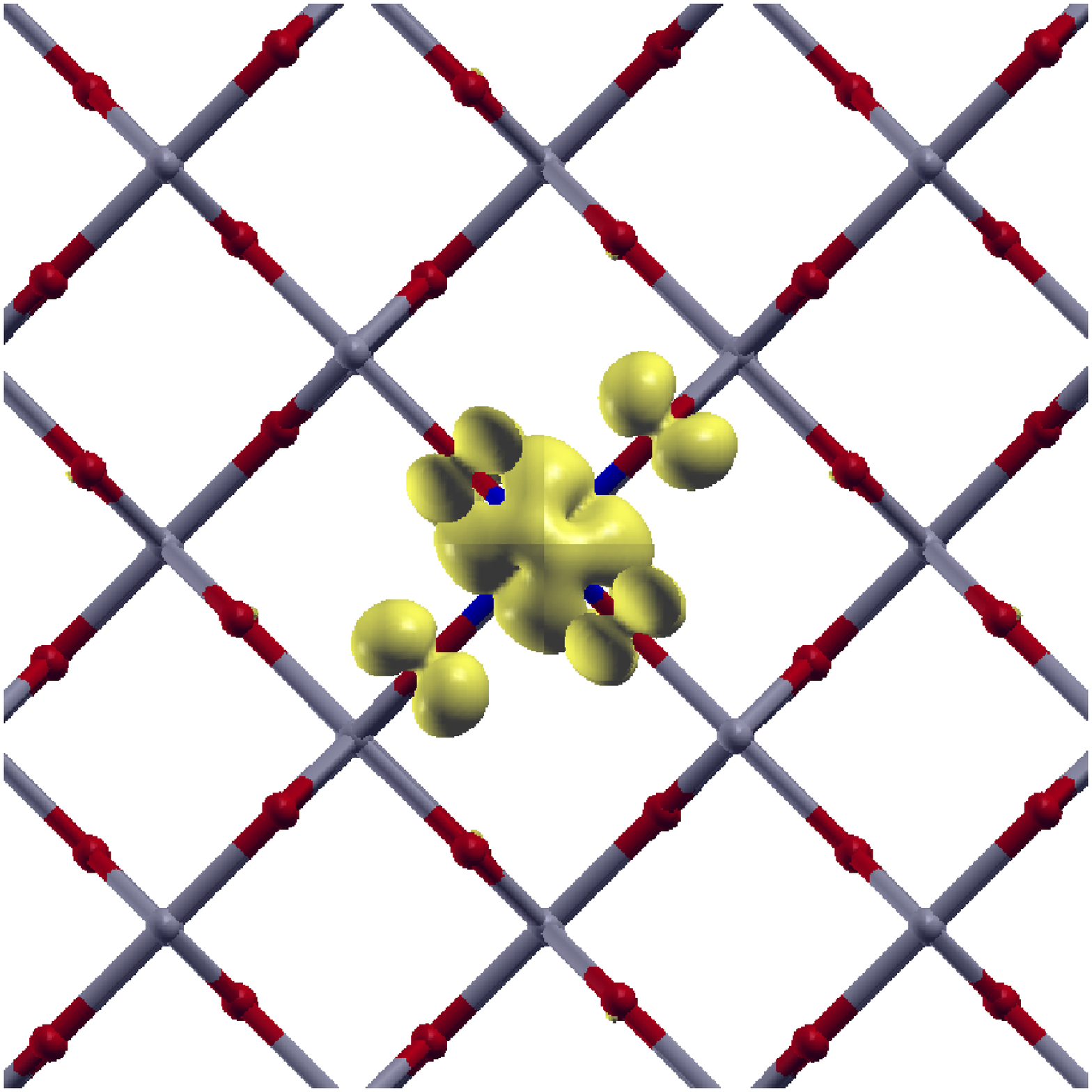}}
  \rput[tl](4,15){\includegraphics[width=9cm]{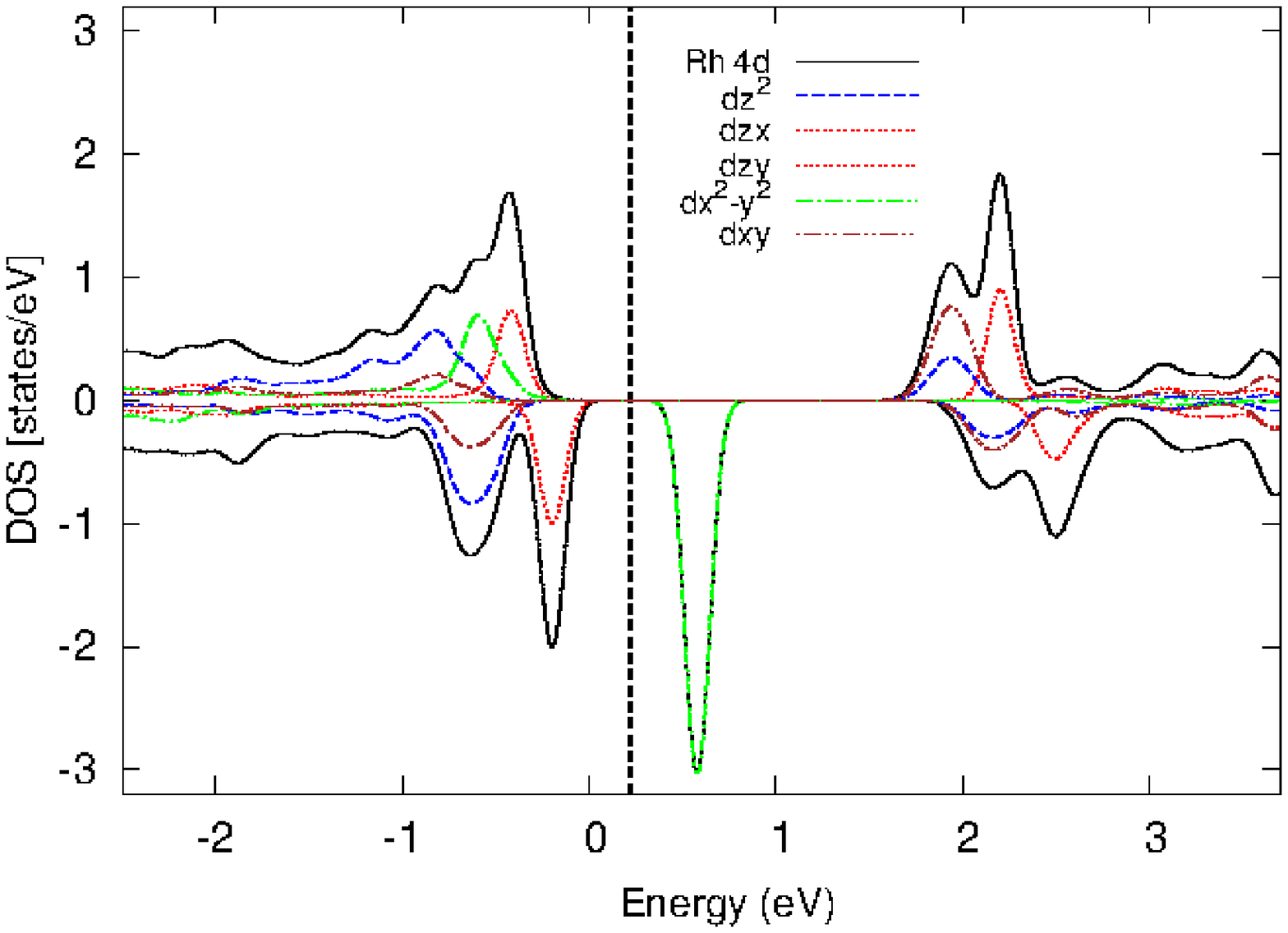}}
\psline[linewidth=0.7pt, linecolor=black]{<->}(4.0,19.2)(4.53,19.2)
\psline[linewidth=0.7pt, linecolor=black]{<->}(4.0,20.8)(6.1,20.8)
\psline[linewidth=0.7pt, linecolor=black]{<->}(4.97,19.2)(6.1,19.2)
\psline[linewidth=0.7pt, linecolor=black]{-}(4.0,21)(4.0,19)
\psline[linewidth=0.7pt, linecolor=black]{-}(4.53,20.15)(4.53,19)
\psline[linewidth=0.7pt, linecolor=black]{-}(4.97,20.15)(4.97,19)
\psline[linewidth=0.7pt, linecolor=black]{-}(6.1,21)(6.1,19)
\rput[tl](3.48,18.8){\scriptsize{0.4 eV}} 
\rput[tl](4.6,21.05){\scriptsize{1.7 eV}}
\rput[tl](5.15,19.0){\scriptsize{1.0 eV}}
\rput[tl](4.8,16.1){\small\bf (a)}
\rput[tl](11.8,16.1){\small\bf (b)}
\rput[tl](8.8,8.1){\small\bf (c)}
\end{pspicture}
\caption{The density of states and projected density of states on the p and d orbitals (a), spin density (b) and projected density of states for the Rh-4d states (c) in Rh-doped rutile model. The zero energy value is set at the top of VB and Fermi energy is represented by the vertical dashed line in (a) and (c).}
 \label{RhPDOS}
\end{figure}

The total DOS with PDOS for the d electrons of Rh and Ti atoms and p electrons of O atoms of the Rh-doped rutile are summarized in Fig.~\ref{RhPDOS}(a). Rh doping increases the asymmetry between the neighbouring atoms and introduces an unpaired d electron from Rh. As a result an intermediate energy band of 0.3 eV width is found approximately 1.0 eV below the CB minima, along with some states at CB and VB tails. This leads to an overall reduction in the band gap by 0.5 eV. To understand the nature of defect states further, consider the electronic configurations of Rh: 4d$^8$5s$^1$ and Ti: 3d$^2$4s$^2$. Substitution of Ti with Rh leads to a formal $+4$ state (Rh$^{4+}$), i.e. 4d$^5$5s$^0$ configuration. Since the stable oxidation state of Rh is +3, an electron get transferred from neighbouring O atoms to Rh$^{4+}$ leading to O$^{1-}$ defects. The band gap states are mainly due to hybridization of Rh 4d and O 2p states. The  spin density plot (Fig.~\ref{RhPDOS}(b)) also shows that the spin densities are strongly localized on the dopant atom and O$^{1-}$ sites in its neighborhood. To further clarify the origin of band gap states, we present in Fig.~\ref{RhPDOS}(c) the five-fold degenerate Rh-4d states. The 2-fold degenerate e$_g$ states split into d$_{x^2-y^2}$ and d$_{z^2}$ states; whereas the 3-fold degenerate t$_{2g}$ states split into doubly degenerate states (d$_{xz}$, d$_{yz}$) and a single state (d$_{xy}$). It is evident from Fig.~\ref{RhPDOS}(c) that only d$_{x^2-y^2}$ state of Rh-4d orbital contributes to band gap state. Rh doping of rutile has also been investigated by Tan et al.~\cite{Tan2012}. Their calculations suggested band gap states very near the VB maxima, whereas in the present case defect states are almost in the middle of band gap. Nonetheless, in both studies, the contribution to band gap states comes from the same  d$_{x^2-y^2}$ state of Rh-4d orbital. The difference in position of band gap states can be attributed to different choice of $U$ parameter. 
 
\par The formation of a VB by orbitals not associated with O 2p but with other elements is indispensable in designing visible-light-driven photocatalysts as it moves top of the VB at a more positive potential than the oxidation potential of H$_2$O to O$_2$ without affecting the CB level and charge migration~\cite{Kudo2007}. Thus an acceptor state at the VB maxima not associated with O 2p but with the Rh$^{+3}$ ions can increase photo activity in Rh-doped TiO$_2$ significantly. Additional states present in the band gap may also contribute to increased photo activity due to electron excitations under visible light irradiation from top of the VB to the band gap states and then from band gap states to the bottom of the CB. Choi et al.~\cite{Choi1994} also suggests that Rh-doped TiO$_2$ shows red shift due to charge transfer between Rh and TiO$_2$ CB or VB; or a d-d transition in the crystal field. These mid-gap states are not without a downside, however, as they can act as recombination centers, consequently leading to a significant reduction in the photocatalytic efficiency.  As per Niishiro et al.~\cite{Niishiro2007}, this recombination is chiefly due to the formation of Rh$^{4+}$. To suppress Rh$^{4+}$, codoping of Rh-doped rutile with high-valent Nb$^{5+}$ could be useful in limiting formation of these recombination centers, and is therefore investigated further in subsequent discussion.
\subsubsection{Nb-doped rutile TiO$_2$}
\begin{figure}
\begin{pspicture}(0,6)(15,24) 
  \rput[tl](0,20){\includegraphics[width=9cm]{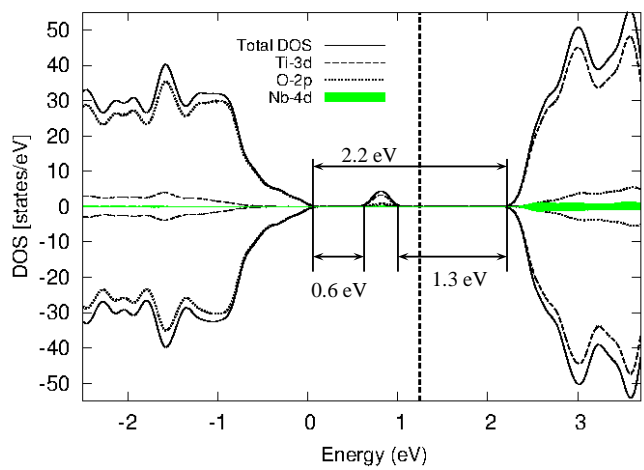}}
  \rput[tl](9,20){\includegraphics[width=6cm]{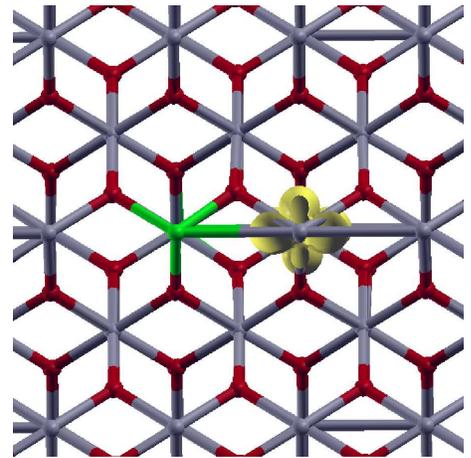}}
\psline[linewidth=0.7pt, linecolor=black]{<->}(4.12,16.4)(4.8,16.4)
\psline[linewidth=0.7pt, linecolor=black]{<->}(4.12,17.6)(6.7,17.6)
\psline[linewidth=0.7pt, linecolor=black]{<->}(5.25,16.4)(6.7,16.4)
\psline[linewidth=0.7pt, linecolor=black]{-}(4.12,17.7)(4.12,16.2)
\psline[linewidth=0.7pt, linecolor=black]{-}(4.8,17.1)(4.8,16.2)
\psline[linewidth=0.7pt, linecolor=black]{-}(5.25,17.1)(5.25,16.2)
\psline[linewidth=0.7pt, linecolor=black]{-}(6.7,17.7)(6.7,16.2)
\rput[tl](4.1,16.05){\scriptsize{0.6 eV}}
\rput[tl](4.5,17.85){\scriptsize{2.2 eV}}
\rput[tl](5.7,16.2){\scriptsize{1.3 eV}}
\rput[tl](4.8,13.1){\small\bf (a)}
\rput[tl](11.8,13.1){\small\bf (b)}
\end{pspicture}
  \caption{The density of states and projected density of states on the p and d orbitals (a) and spin density plot (b) for one Nb atom doped in 72-atoms rutile model. The zero energy value is set at the top of VB and Fermi energy is represented by the vertical dashed line in (a).}
  \label{NbPDOS}
\end{figure}

Before proceeding to charge compensated doping, let us describe observations on Nb-monodoped rutile. From the total DOS, PDOS and accompanying spin density plots shown in Fig.~\ref{NbPDOS}, it can be deduced that the substitution of Ti$^{4+}$ atom with Nb$^{5+}$ in TiO$_2$ results in a narrow peak of 0.3 eV width in the band gap approximately 1.3 eV below the CB edge. It is noted here that our independent calculations using DFT alone did not show a band gap state with excess charge occupying the bottom of the CB. This implies the necessity and significance of the DFT$+U$ approach in reproducing electronic states and band structure commensurate with the experimental observations in an accurate and physically consistent manner. The PDOS plot (Fig.~\ref{NbPDOS}(a)) also suggests that the defect state is mainly due to contribution from Ti atoms. Experimental data have also shown a defect state in the band gap well separated from the CB ~\cite{Morris2000}, with a height proportional to the Nb concentration. This observations on Nb-doped TiO$_2$ with DFT+$U$ agrees with previous theoretical investigations by Morgan et al.~\cite{Morgan2009b}. Considering the electronic configuration of Nb: 4p$^6$4d$^3$5s$^2$, substitution of Ti by Nb should result into Nb$^{4+}$ state, i.e. 4p$^6$4d$^1$5s$^0$ configuration. To obtain a stable oxidation state of +5 the extra electron from Nb$^{4+}$ is preferentially transferred to Ti$^{4+}$ and therefore Ti$^{3+}$ state is formed. This is confirmed by spin density plot (Fig~\ref{NbPDOS}(b)) which shows that the spin density strongly localized on a single Ti site in a nearest-neighbour position relative to the Nb dopant.
\subsubsection{(Rh,Nb)- codoped rutile TiO$_2$}
\par The presence of Rh states at VB maxima and intermediate band with considerable bandwidth in Rh-doped rutile indicates great potential of using the Rh-doped material for visible light photocatalytic applications. This intermediate band offers a stepping stone for the absorption of low energy photons via the excitation of electrons at the VB maximum to the intermediate bands, from where they can be excited again above the CB for effective photoactivity of TiO$_2$. Nonetheless, these band gap states can also act as recombination centers, see e.g. experimental observations by Niishiro et al.~\cite{Niishiro2007}. This charge recombination renders Rh-doped TiO$_2$ as ineffective photocatalyst. These recombination centers need to be eliminated by suppressing Rh$^{4+}$ ions. This can be achieved by codoping of TiO$_2$ by a combination of p-type and n-type dopants, as also suggested by ~\cite{Niishiro2007,Long2011}. In this context, further codoping of Rh-doped rutile with Nb was investigated with the goal of eliminating of Rh$^{4+}$ ions through charge compensation. For studying codoped system, we constructed 6 nearest neighbour codoped structures. These structure were relaxed and the structure having Rh-Nb distance of 2.95 $\AA$ showed lowest configuration energy and is therefore considered for further analysis.
\par
\begin{figure}
\begin{pspicture}(0,6)(15,24) 
  \rput[tl](0,23){\includegraphics[width=9cm]{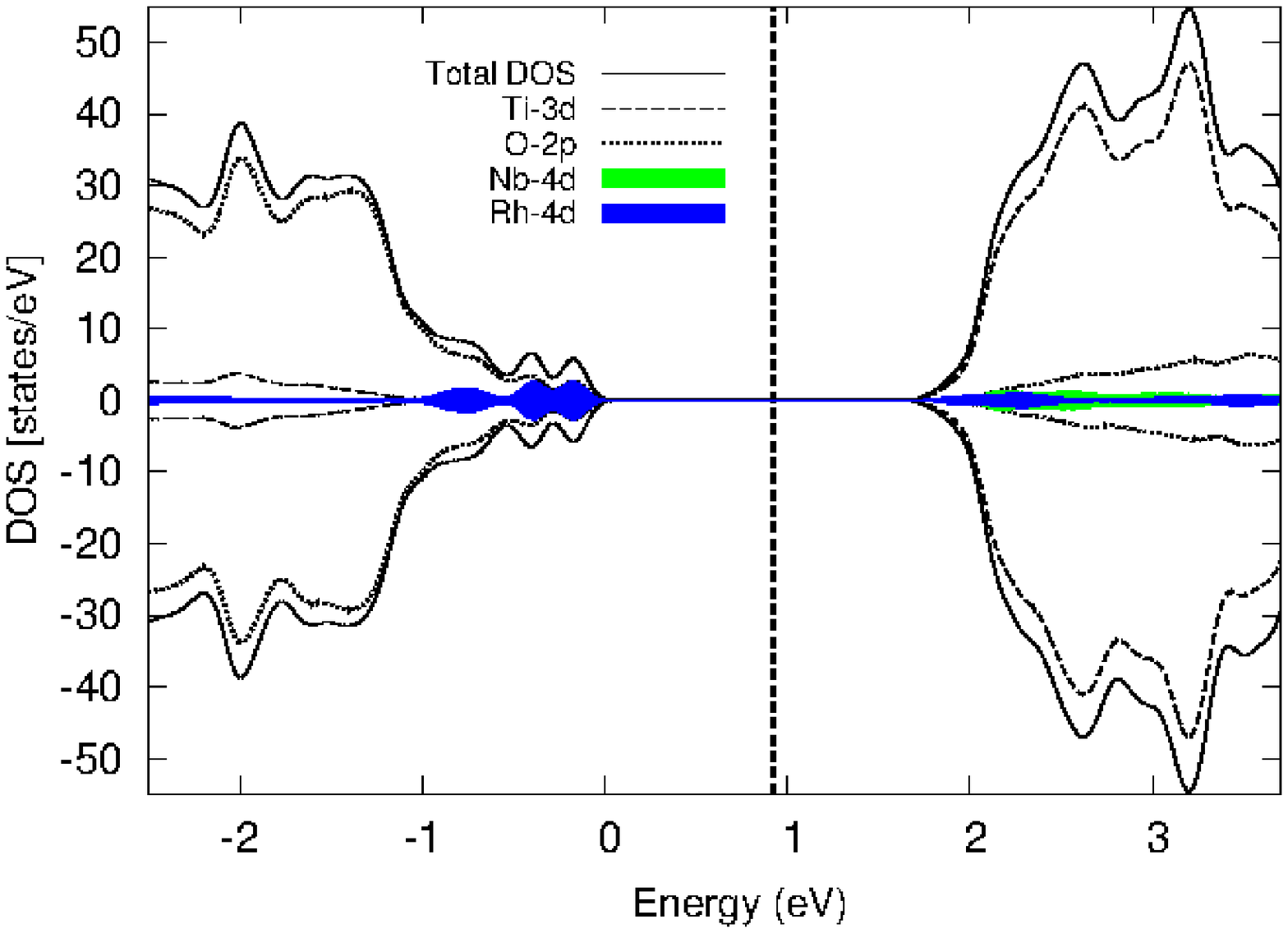}}
  \rput[tl](9,23){\includegraphics[width=6cm]{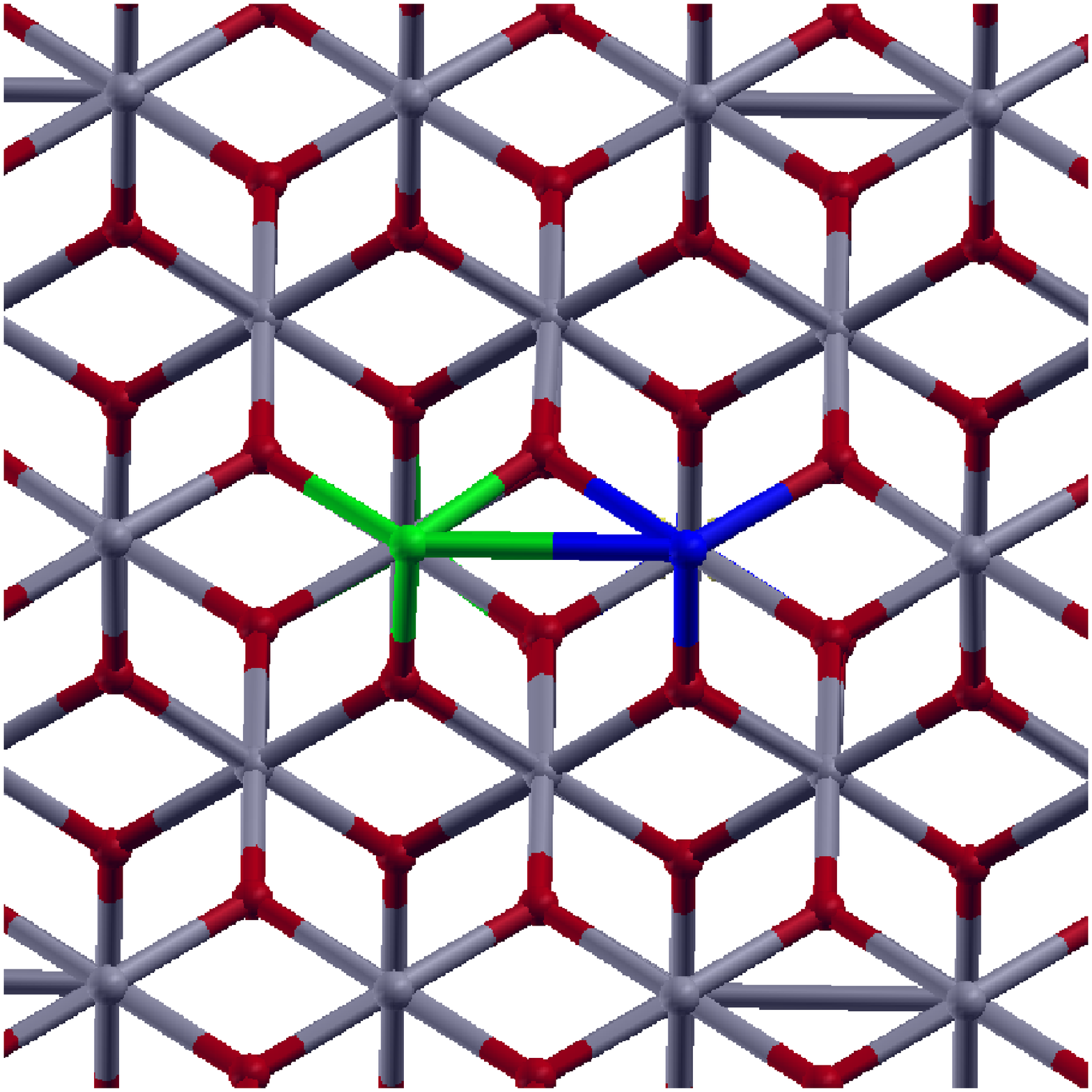}}
  \rput[tl](4,15){\includegraphics[width=9cm]{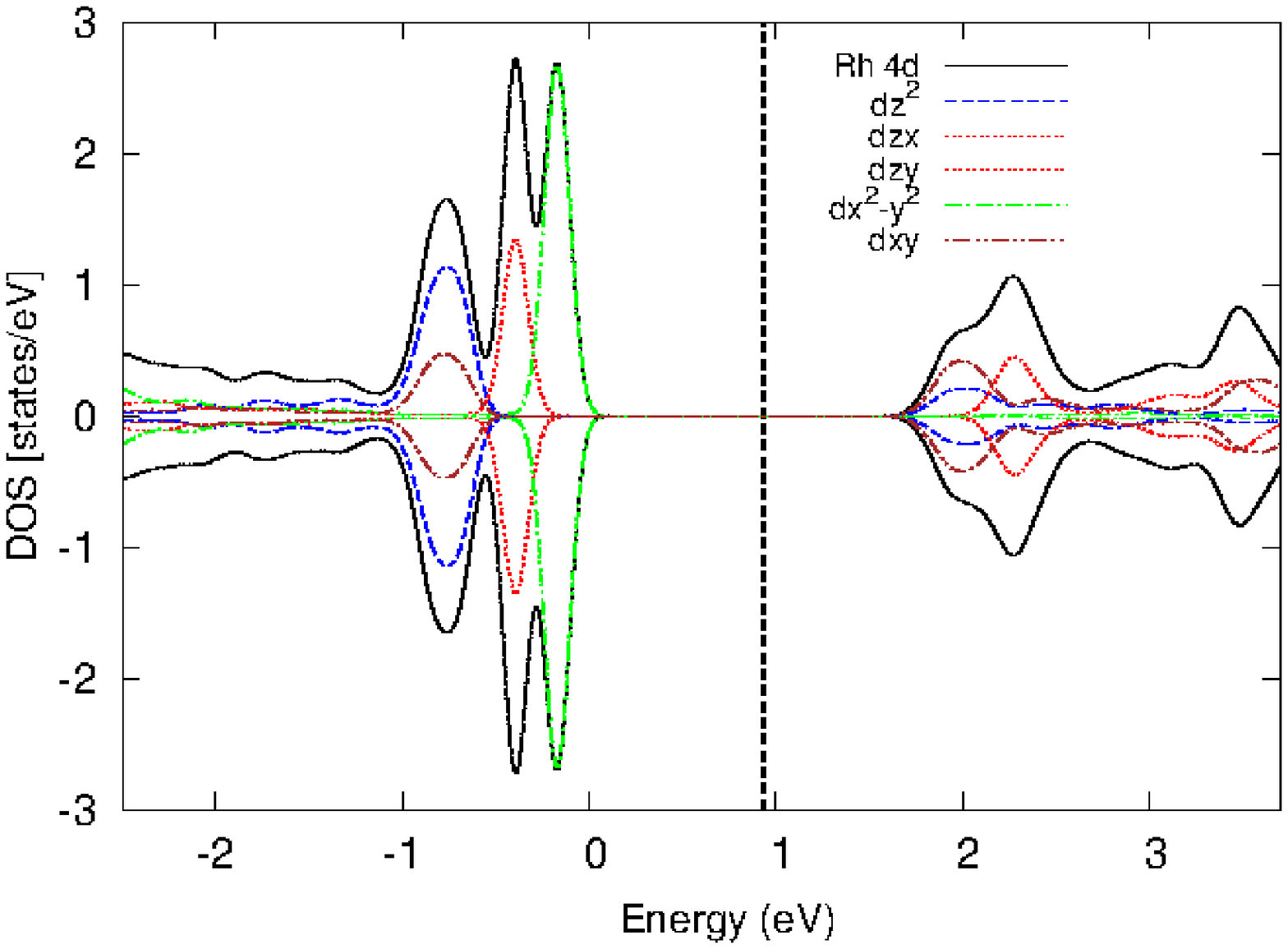}}
\psline[linewidth=0.7pt, linecolor=black]{<->}(4.05,19.2)(6.1,19.2)
\psline[linewidth=0.7pt, linecolor=black]{-}(4.05,19)(4.05,20.1)
\psline[linewidth=0.7pt, linecolor=black]{-}(6.1,19)(6.1,20.1)
\rput[tl](4.3,18.95){\scriptsize{1.7 eV}}
\rput[tl](4.8,16.1){\small\bf (a)}
\rput[tl](11.8,16.1){\small\bf (b)}
\rput[tl](8.8,8.1){\small\bf (c)}
\end{pspicture}
\caption{The density of states and projected density of states on the p and d orbitals (a), spin density  (b) and projected density of states for the Rh-4d states (c) in (Rh,Nb)- codoped rutile model. The zero energy value is set at the top of VB and Fermi energy is represented by the vertical dashed line in (a) and (c).}
  \label{RhNbPDOS}
\end{figure}

 Total spin polarized DOS and PDOS for the d electrons of Rh, Nb and Ti atoms and p electrons of O atoms of the (Rh,Nb)- codoped rutile are shown in Fig.~\ref{RhNbPDOS}(a). Based on the analyses of Rh and Nb monodoping described earlier in the paper, Rh dopant requires an extra electron to achieve the stable +3 oxidation state whereas Nb dopant introduces an extra electron into the system. Thus, when rutile is codoped with Rh and Nb the extra electron on Ti$^{3+}$ due to Nb transfers to Rh dopant such that the Rh species attains a +3 state (4d$^5$5s$^1$) while the Nb species attains a +5 (4p$^6$4d$^0$5s$^0$) state, with both resulting states being more stable. This charge compensation of (Rh,Nb)- codoped system can also be inferred in the spin density plot (Fig~\ref{RhNbPDOS}(b)) by the absence of excess spin density states. This codoping also reduces the band gap by 0.5 eV relative to undoped TiO$_2$, due to the new VB consisting of hybridized Rh and O 2p orbitals which results in decreased band gap energy without affecting the CB level. The Rh d$_{x^2-y^2}$ orbital, which was earlier responsible for band gap states in Rh-doped rutile, now shifts to a fully filled state at valence band minimum, as shown in Fig.~\ref{RhNbPDOS}(c). Furthermore, the codoped structure does not require formation of O vacancies to provide extra electron to Rh. This, in turn leads to a smaller degree of defectiveness and elimination of recombination centers. Our analysis thus establishes the beneficial aspects of charge compensated codoping and seems to explain findings of Niishiro et al.~\cite{Niishiro2007} for (Rh,Nb)- codoped TiO$_2$.
\subsection{Charge density}
\begin{figure*}[]
 \centerline{{\hbox{ 
   \epsfxsize=2.50in
   \epsffile{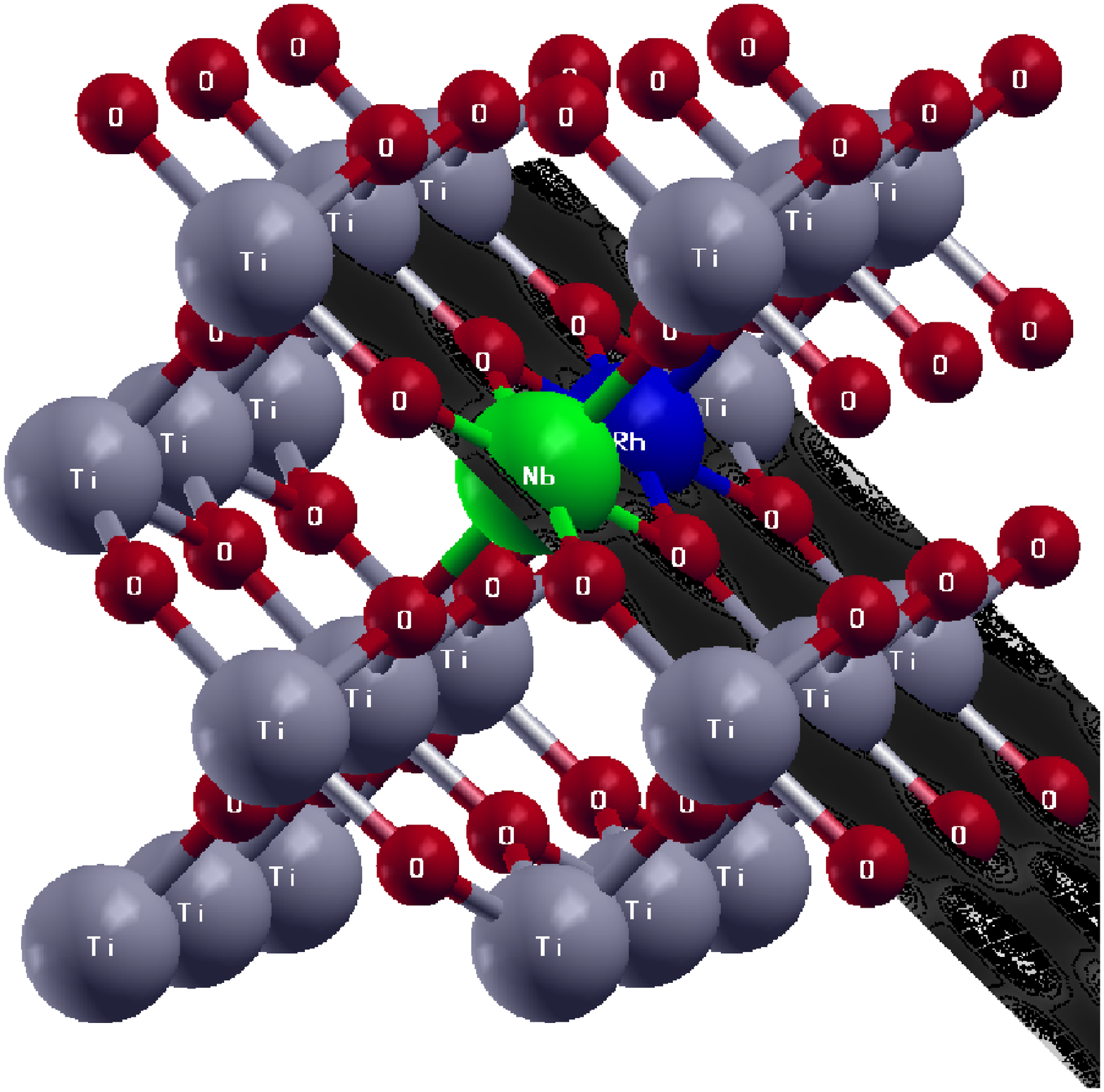}}}
 {{\hbox{
   \epsfxsize=2.50in
   \epsffile{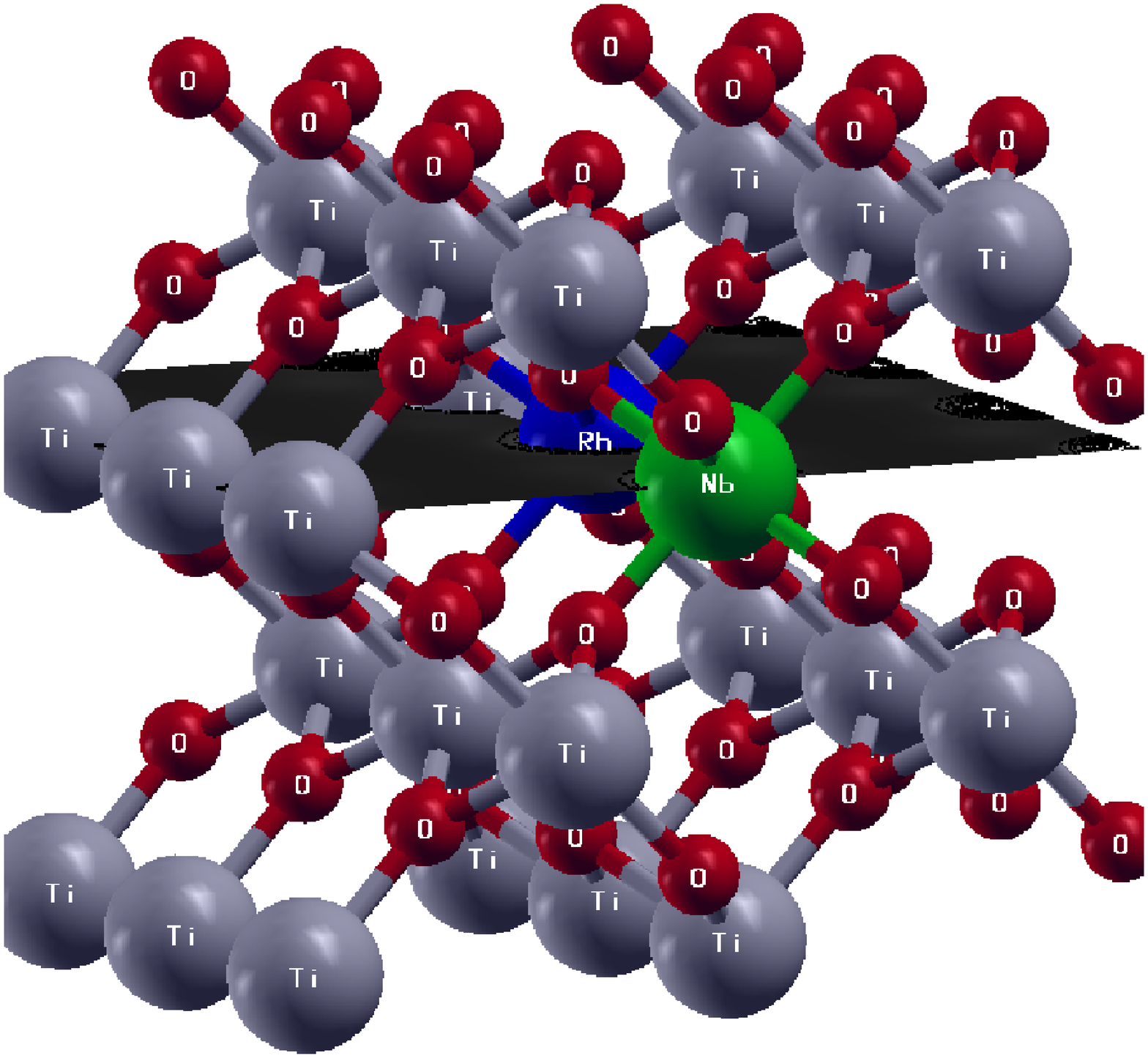}}}
  }}
\hbox{\hspace{1.40in} (a) \hspace{2.40in} (b)} 
 \centerline{{\hbox{ 
   \epsfxsize=2.50in
   \epsffile{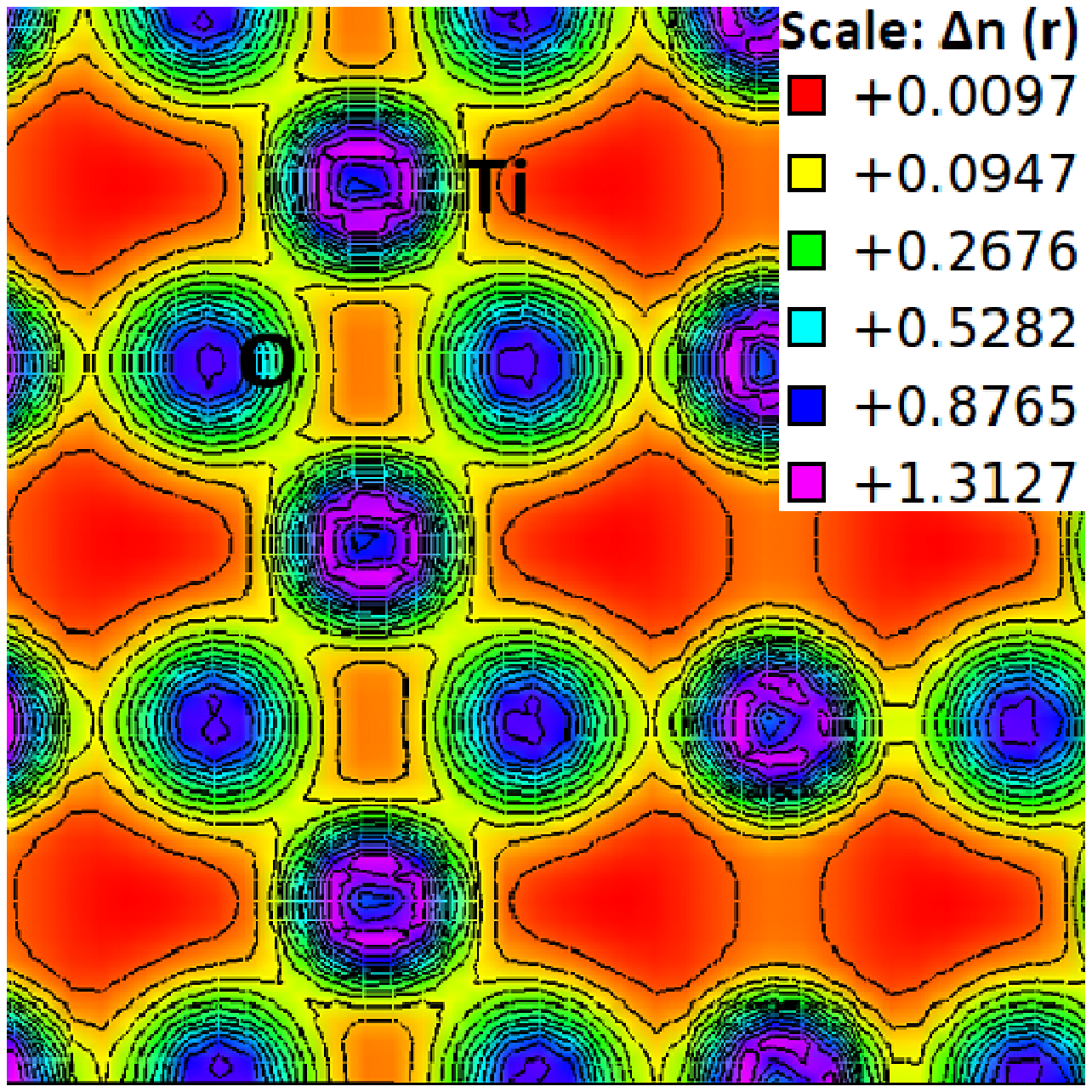}}}
 {{\hbox{
   \epsfxsize=2.50in
   \epsffile{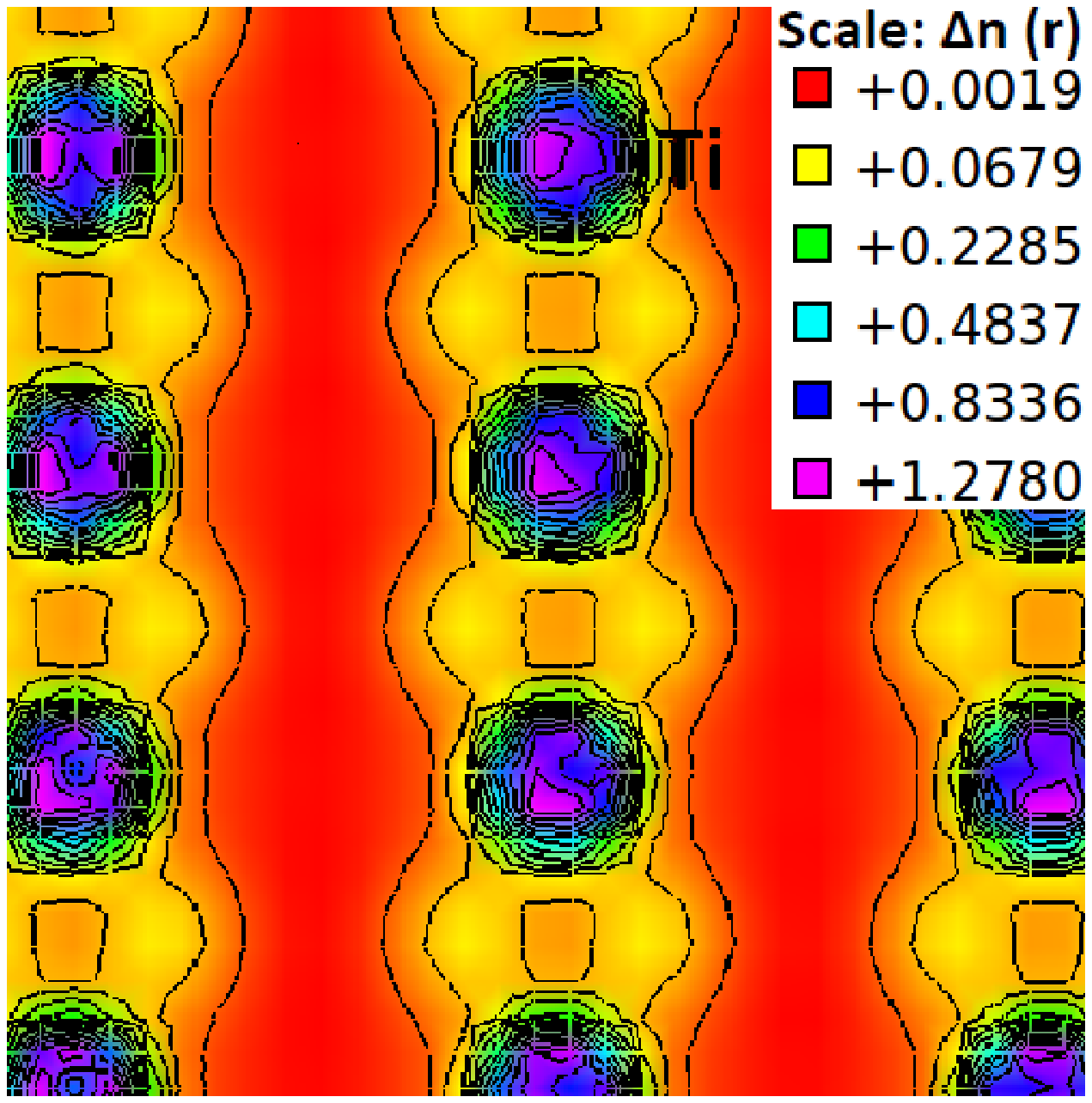}}}
  }}
\hbox{\hspace{1.40in} (c) \hspace{2.40in} (d)} 
 \centerline{{\hbox{ 
   \epsfxsize=2.50in
   \epsffile{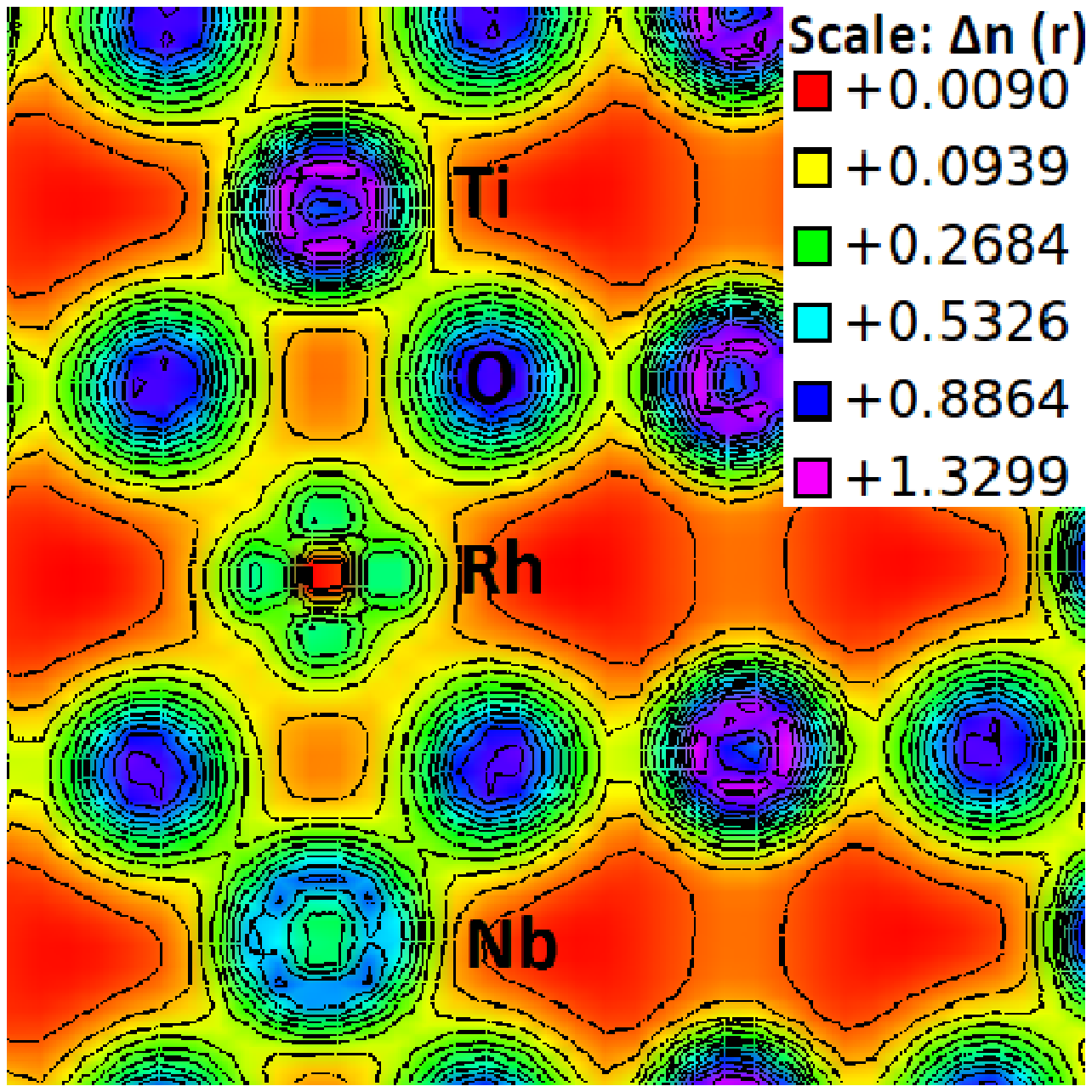}}}
{{\hbox{
   \epsfxsize=2.50in
   \epsffile{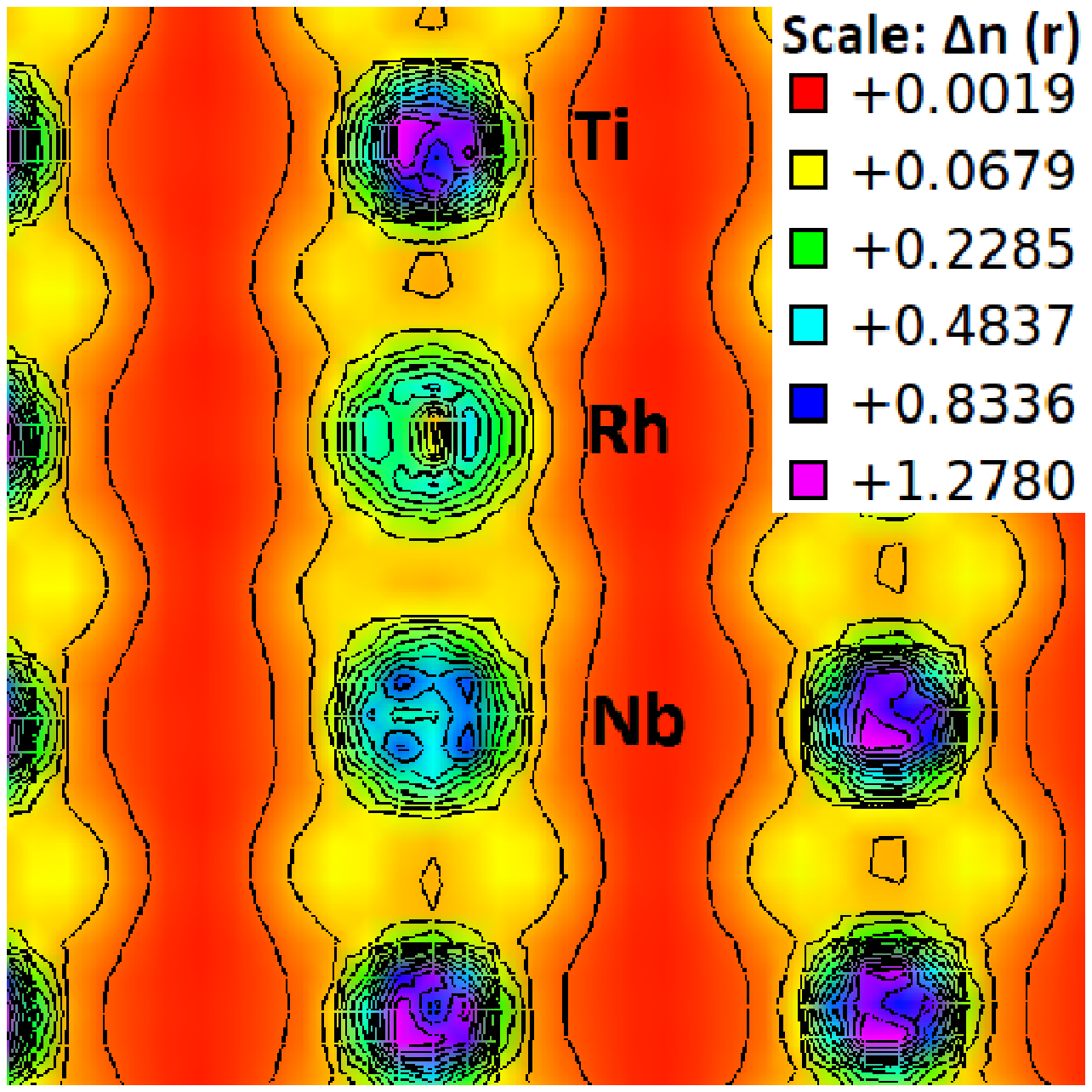}}}
  }}
\hbox{\hspace{1.40in} (e) \hspace{2.40in} (f)} 
\caption{Two dimensional 110 (a) and 010 (b) planes indicated in black for which charge density contour plots (electron bohr$^{-3}$, (1bohr= 0.529$\AA$) units) are depicted. The charge density plot for undoped rutile in 110 plane (a), undoped rutile in 010 plane (b), (Rh,Nb)- codoped rutile in 110 plane (c) and (Rh,Nb)- codoped rutile in 010 plane (d). The red regions denote the area with nearly zero electron density and the pink regions denote the highest electron density.}
 \label{CD}
\end{figure*}

To substantiate the electronic properties further and to study the variation of chemical bonding induced by Rh and Nb codoping, the electron charge density distributions for undoped and charge compensated systems were also analyzed. As depicted in Fig.~\ref{CD}, the contour plots of the charge density on (1 1 0) and (0 1 0) represent the effect of dopants on nearest neighbouring O and Ti atoms, respectively. The charge density around Rh and Nb atoms is nearly spherically distributed with a slight deformation toward their nearest-neighboring atoms. Fig.~\ref{CD}(e) shows that there are clear covalent bridges between dopants and O atoms, which strengthens the judgement of certain covalent character of (Rh,Nb)- codoped rutile. Furthermore, it can be seen that the overlap of Nb 4d orbital with O 2p is greater than that of Rh 4d orbital with O 2p, which indicates that the degree of covalency between Nb and O is stronger than that between Rh and O. A weaker interaction between Rh and O atoms as compared to that between Ti and O atoms makes the Rh-O bond longer and leads to smaller binding energy. Consequently the O 2p states, which are dominating in valence band edge, move toward high energy levels and shift the VB maximum to increased energy, thereby reducing the band gap. From (0 1 0) plane, on comparing with undoped rutile (Fig.~\ref{CD}(d)), it can be seen that there is a considerable charge accumulation in the bonding regions of Rh and Nb dopants (Fig.~\ref{CD}(f)). This confirms the sharing (or transfer) of electron between the dopants. The accumulation of charge in the region close to the dopants also suggests the formation of a metallic contact. It should be  noted that the electron charge density plotted on plane next to doped elements is almost completely unaffected by the presence of the doping elements.
\section{Conclusions}
In this work, ab-initio calculations were conducted to understand conflicting experimental results for photocatalytic behavior of Rh-doped TiO$_2$ and to explore a more efficient photocatalysts by codoping it with Nb. DFT+$U$ methodology was utilized to remove excessive delocalization of Ti d-orbitals present in standard DFT calculations. Doping rutile with Rh atom induces states at valence band minima and an intermediate band that leads to a red shift in optical absorption edge; also observed in previous experimental investigations. But since these states may also act as recombination centers due to presence of Rh$^{4+}$ ions, codoping with Nb was investigated. Following conclusions can be derived from the present study:
\begin{enumerate}
\item  The formation energy of Rh-doped system reduces appreciably on codoping it with Nb, suggesting a spontaneous formation of (Rh,Nb)- codoped rutile TiO$_2$ at room temperature.
\item Codoping affects the symmetry, leading to a weaker interaction and longer bond between Rh and O as compared to Nb and O. This structural distortion reduces the binding energy and the O 2p states tend to move from VB edge towards high energy levels, reducing the overall band gap. The hybridized O 2p and Rh 4d states at the top of VB in codoped rutile are thought to be responsible for the red shift of optical absorption edge, confirming previous experimental studies~\cite{Niishiro2007,Choi1994,Kitano2011}. 
\item Rh dopant requires an extra electron to achieve the stable +3 oxidation state whereas Nb dopant introduces an extra electron into the system. Thus, when the system is codoped with Rh and Nb, both dopants attain their stable oxidation states. It means that the creation of oxygen vacancy is not required in a codoped structure. This leads to a significant reduction in the level of defectiveness and thus an enhanced photocatalytic activity.
 \item Unlike Rh-doped rutile that has an isolated band gap state which may act as a recombination center, charge compensated (Rh,Nb)-codoping does not show such band gap states which commensurate with experimental findings of Niishiro et al.~\cite{Niishiro2007}. This elimination of recombination centers in combination with a band gap reduction by 0.5 eV makes codoped rutile a better photocatalyst.
\end{enumerate}
Overall, the present study reinforces the usefulness of charge compensated codoping for reducing the band gap in photocatalyts, without creation of unwanted recombination centers. In the near future, we plan to apply this technique further to amorphous TiO$_2$ in the hope of developing more efficient and cost-effective visible light photocatalysts.
\section{Acknowledgments}
Computations were performed on the HPC supercomputer at the SciNet HPC
Consortium~\cite{Loken2010_scinet} and Calcul Quebec/Compute Canada. SciNet is funded by: the Canada Foundation for Innovation under the auspices of Compute Canada; the Government of Ontario; Ontario Research Fund - Research Excellence; and the University of Toronto. The authors gratefully acknowledge the continued support of above organizations.
\newpage
\clearpage
\bibliography{Bibtex-RhNbDoped-TiO2}

\begin{thebibliography}{10}

\bibitem{Diebold2011}
U.~Diebold.
\newblock Photocatalysts closing the gap.
\newblock {\em Nat Chem}, 3:271, 2011.

\bibitem{Fujishima2008}
A.~Fujishima, X.~Zhang, and D.~A. Tryk.
\newblock Tio(2) photocatalysis and related surface phenomena.
\newblock {\em Sur Sci Rep}, 63:515, 2008.

\bibitem{Ni2007}
M.~Ni, M.~K.~H. Leung, D.~Y.~C. Leung, and K.~Sumathy.
\newblock A review and recent developments in photocatalytic water-splitting
  using tio2 for hydrogen production.
\newblock {\em Rene Sust Energ Rev}, 11:401, 2007.

\bibitem{Ollis1993}
D.~F. Ollis and H.~Al-Ekabi.
\newblock {\em Photocatalytic purification and treatment of water and air}.
\newblock Amsterdam: Elsevier Science, 1993.

\bibitem{Wang2001}
Y.-Q. Wang, S.-G. Chen, X.-H. Tang, O.~Palchik, A.~Zaban, Y.~Koltypin, and
  A.~Gedanken.
\newblock Mesoporous titanium dioxide: sonochemical synthesis and application
  in dye-sensitized solar cells.
\newblock {\em J Mater Chem}, 11:521, 2001.

\bibitem{Xu2002}
A.~W. Xu, Y.~Gao, and H.~Q. Liu.
\newblock The preparation, characterization, and their photocatalytic
  activities of rare-earth-doped tio2 nanoparticles.
\newblock {\em J Catal}, 207:151, 2002.

\bibitem{Diebold2003}
U.~Diebold.
\newblock The surface science of titanium dioxide.
\newblock {\em Surface Science Reports}, 48(5-8):53--229, 2003.

\bibitem{Yang2007}
K.~Yang, Y.~Dai, and B.~Huang.
\newblock Understanding photocatalytic activity of s- and p-doped tio(2) under
  visible light from first-principles.
\newblock {\em J Phys Chem C}, 111:18985, 2007.

\bibitem{Chen2011a}
X.~Chen, L.~Liu, P.~Y. Yu, and S.~S. Mao.
\newblock Increasing solar absorption for photocatalysis with black
  hydrogenated titanium dioxide nanocrystals.
\newblock {\em Science}, 331:746, 2011.

\bibitem{Fuerte2001}
A.~Fuerte, M.~D. Hernandez-Alonso, A.~J. Maira, A.~Martinez-Arias,
  M.~Fernandez-Garcia, J.~C. Conesa, and Soria.
\newblock Visible light-activated nanosized doped-tio2 photocatalysts.
\newblock {\em Chem Commun}, page 2718, 2001.

\bibitem{Choi1994}
W.~Y. Choi, A.~Termin, and M.~R. Hoffmann.
\newblock The role of metal-ion dopants in quantum-sized tio2 - correlation
  between photoreactivity and charge-carrier recombination dynamics.
\newblock {\em J Phys Chem}, 98:13669, 1994.

\bibitem{Gao2005}
L.~Gao, H.~Liu, and S.~Jing.
\newblock The preparation of rutile tio2 doped with rare earth metal ions,
  sulfur and their photocatalytic properties.
\newblock {\em Materials Science Forum}, 486-487:53, 2005.

\bibitem{Kitano2011}
S.~Kitano, K.~Hashimoto, and H.~Kominami.
\newblock Photocatalytic degradation of 2-propanol over metal-ion-loaded
  titanium(iv) oxide under visible light irradiation: Effect of physical
  properties of nano-crystalline titanium(iv) oxide.
\newblock {\em Appl Catal B-Environ}, 101:206, 2011.

\bibitem{Niishiro2007}
R.~Niishiro, R.~Konta, H.~Kato, W.-J. Chun, K.~Asakura, and A.~Kudo.
\newblock Photocatalytic o2 evolution of rhodium and antimony-codoped
  rutile-type tio2 under visible light irradiation.
\newblock {\em J Phys Chem C}, 111:17420, 2007.

\bibitem{QE}
P.~Giannozzi, S.~Baroni, N.~Bonini, M.~Calandra, R.~Car, C.~Cavazzoni,
  D.~Ceresoli, G.L. Chiarotti, M.~Cococcioni, I.~Dabo, et~al.
\newblock Quantum espresso: a modular and open-source software project for
  quantum simulations of materials.
\newblock {\em J Phys: Condens Matter}, 21:395502, 2009.

\bibitem{Perdew1996_pbe}
J.P. Perdew, K.~Burke, and M.~Ernzerhof.
\newblock Generalized gradient approximation made simple.
\newblock {\em Phys Rev Lett}, 77:3865, 1996.

\bibitem{Vanderbilt1990_PRB}
D.~Vanderbilt.
\newblock Soft self-consistent pseudopotentials in a generalized eigenvalue
  formalism.
\newblock {\em Phys Rev B}, 41:7892, 1990.

\bibitem{Monkhorst1976}
H.~J. Monkhorst and J.~D. Pack.
\newblock Special points for brillouin-zone integrations.
\newblock {\em Phys Rev B}, 13:5188, 1976.

\bibitem{ICSD2010}
Inorganic crystal structure database (icsd).
\newblock {\em NIST Release}, 2010.

\bibitem{Zeng2010}
Z.~L. Zeng, G.~Zheng, X.~C. Wang, K.~H. He, Q.~L. Chen, L.~Yu, and Q.~B. Wang.
\newblock First-principles study on the structural and electronic properties of
  double n atoms doped-rutile tio(2).
\newblock {\em J Atom Mol Sci}, 1:177, 2010.

\bibitem{Shao2008}
G.~Shao.
\newblock Electronic structures of manganese-doped rutile tio(2) from first
  principles.
\newblock {\em J Phys Chem C}, 112:18677, 2008.

\bibitem{Iwaszuk2011}
A.~Iwaszuk and M.~Nolan.
\newblock Charge compensation in trivalent cation doped bulk rutile tio(2).
\newblock {\em J Phys-Condens Matter}, 23:334207, 2011.

\bibitem{Deskins2007}
N.~A. Deskins and M.~Dupuis.
\newblock Electron transport via polaron hopping in bulk tio(2): A density
  functional theory characterization.
\newblock {\em Phys Rev B}, 75:195212, 2007.

\bibitem{Wendt2008}
S.~Wendt, P.~T. Sprunger, E.~Lira, G.~K.~H. Madsen, Z.~Li, J.~O. Hansen,
  J.~Matthiesen, A.~Blekinge-Rasmussen, E.~Laegsgaard, B.~Hammer, and
  F.~Besenbacher.
\newblock The role of interstitial sites in the ti3d defect state in the band
  gap of titania.
\newblock {\em Science}, 320:1755, 2008.

\bibitem{Thomas2007}
A.~G. Thomas, W.~R. Flavell, A.~K. Mallick, A.~R. Kumarasinghe, D.~Tsoutsou,
  N.~Khan, C.~Chatwin, S.~Rayner, G.~C. Smith, R.~L. Stockbauer, S.~Warren,
  T.~K. Johal, S.~Patel, D.~Holland, A.~Taleb, and F.~Wiame.
\newblock Comparison of the electronic structure of anatase and rutile tio2
  single-crystal surfaces using resonant photoemission and x-ray absorption
  spectroscopy.
\newblock {\em Phys Rev B}, 75:035105, 2007.

\bibitem{Moller2010}
J.~S. Moller, H.~K. Henrik, B.~Hinnemann, G.~K. H.~Madsen Madsen, and
  B.~Hammer.
\newblock Dft + u study of defects in bulk rutile tio2.
\newblock {\em J Chem Phys}, 133:144708, 2010.

\bibitem{Morgan2007}
B.~J. Morgan and G.~W. Watson.
\newblock A dft + u description of oxygen vacancies at the tio2 rutile (1 1 0)
  surface.
\newblock {\em Surf Sci}, 601:5034, 2007.

\bibitem{Yang2010}
S.~Yang and L.~E. Halliburton.
\newblock Fluorine donors and ti3+ ions in tio(2) crystals.
\newblock {\em Phys Rev B}, 81:035204, 2010.

\bibitem{Shannon1976}
R.D. Shannon.
\newblock Revised effective ionic radii and systematic studies of interatomie
  distances in halides and chaleogenides.
\newblock {\em Acta Crystallogr}, A32:751, 1976.

\bibitem{Nolan2011}
M.~Nolan.
\newblock Charge compensation and ce3+ formation in trivalent doping of the
  ceo2(110) surface: The key role of dopant ionic radius.
\newblock {\em Phys Chem}, 115:6671, 2011.

\bibitem{Tan2012}
Z.Y. Tan, L.L. Wang, Y.C. Yang, and W.Z. Xiao.
\newblock Half-metallicity in rh-doped tio2 from ab initio calculations.
\newblock {\em Eur. Phys. J. B}, 85 (4):138, 2012.

\bibitem{Kudo2007}
A.~Kudo.
\newblock Photocatalysis and solar hydrogen production.
\newblock {\em Pure Appl Chem}, 79(11):1917, 2007.

\bibitem{Morris2000}
D.~Morris, Y.~Dou, J.~Rebane, C.~E. Mitchell, R.~G. Egdell, D.~S.~L. Law,
  A.~Vittadini, and M.~Casarin.
\newblock Photoemission and stm study of electronic structure of nb-doped tio2.
\newblock {\em Phys Rev B}, 61:13445, 2000.

\bibitem{Morgan2009b}
B.~J. Morgan, D.~O. Scanlon, and G.~W. Watson.
\newblock Small polarons in nb- and ta-doped rutile and anatase tio2.
\newblock {\em J Mater Chem}, 19:5175, 2009.

\bibitem{Long2011}
R.~Long and N.~J. English.
\newblock Tailoring the electronic structure of tio(2) by cation codoping from
  hybrid density functional theory calculations.
\newblock {\em Phys Rev B}, 83:155209, 2011.

\bibitem{Loken2010_scinet}
C.~Loken, D.~Gruner, L.~Groer, R.~Peltier, N.~Bunn, M.~Craig, T.~Henriques,
  J.~Dempsey, C.-H. Yu, J.~Chen, L.~J. Dursi1, J.~Chong, S.~Northrup,
  J.~Pinto1, N.~Knecht, and R.~V. Zon.
\newblock Scinet: Lessons learned from building a power-efficient top-20 system
  and data centre.
\newblock {\em J Phy: Conf Ser}, 256:012026, 2010.

\end{thebibliography}
\bibliographystyle{unsrt}
\newpage
\clearpage
\newpage
\clearpage
\newpage
\clearpage
\newpage
\clearpage
\newpage
\clearpage
\end{document}